\documentclass{aa}
\usepackage[colorlinks]{hyperref}
\usepackage{microtype}
\usepackage{natbib}
\usepackage{graphicx}
\usepackage{xspace}

\usepackage{txfonts}
\newcommand{\src}{GRO~J1744$-$28\xspace}
\newcommand{\xmm}{\emph{XMM-Newton}\xspace}

\bibpunct{(}{)}{;}{a}{}{,} 

\begin{document} 

   \title{Observations of \src in quiescence with \xmm}
   \author{V. Doroshenko
          \inst{1,2}
          \and
          V. Suleimanov\inst{1,3,2}
          \and
          S.~Tsygankov\inst{4,2}
          \and
          J. M\"onkk\"onen\inst{4}
          \and
          L. Ji\inst{1}
          \and
          A.~Santangelo\inst{1}}

   \institute{
   Institut f\"ur Astronomie und Astrophysik, Universit\"at T\"ubingen, Sand 1, D-72076 T\"ubingen, Germany
       \and
             Space Research Institute of the Russian Academy of Sciences, Profsoyuznaya Str. 84/32, Moscow 117997, Russia
       \and
             Kazan Federal University, Kremlevskaya str. 18, Kazan 42008, Russia
       \and
       Department of Physics and Astronomy, FI-20014 University of Turku, Finland}

\abstract
{We report on the deep observations of the ``bursting pulsar'' \src,  which were performed
 with \xmm and aimed to clarify the origin of its X-ray emission in quiescence. We detect the source at a luminosity level of
$\sim10^{34}$\,erg\,s$^{-1}$ with an X-ray
spectrum that is consistent with the power law, blackbody, or accretion-heated neutron star atmosphere models. The improved X-ray localization of the source allowed us to confirm the
previously identified candidate optical counterpart as a relatively massive G/K~III star
at 8\,kpc close to the Galactic center,
implying an almost face-on view of the binary system. Although we could only find a nonrestricting upper limit on the pulsed fraction of $\sim20$\%, the observed hard X-ray
spectrum and strong long-term variability of the X-ray flux suggest that the source
is also still accreting when not in outburst. The luminosity corresponding to the onset of centrifugal inhibition of accretion is thus estimated to be at least two orders of magnitude lower than previously reported. We discuss this finding in the context of previous studies and argue that the results indicate a multipole structure in the magnetic field with the first dipole term of  $\sim 10^{10}$\,G, which is much lower than previously assumed.}


\maketitle
%

\section{Introduction} The bright hard X-ray transient \src was discovered near the
Galactic center with the Burst and Transient Source Experiment on the Compton
Gamma-Ray Observatory (CGRO BATSE) during its 1995 outburst. The source is bursting and pulsating with a spin period of 467\,ms
\citep{1995IAUC.6272....1F,1996IAUC.6285....1F,1996IAUC.6286....1K}. The bursting activity makes \src a very peculiar X-ray pulsar, usually referred to as a ``bursting pulsar''. 
The origin of the bursts is not fully understood. However, as largely discussed in literature, bursts could arise from instabilities in the accretion flow at the inner edge of the accretion disk, which could produce fluctuations in the accretion rate \citep{1978Natur.271..630H}. This hypothesis is strongly supported by the decrease in the accretion rate immediately following the bursts. On the other hand, a thermal component in the burst X-ray spectra has been observed as well
\citep{2015MNRAS.452.2490D}, which suggests that bursts may be accompanied by thermonuclear burning.

An analysis of the spin modulation of the pulsar allowed us to constrain the orbital
parameters of the system and put an upper limit on the companion mass and on the
magnetic field of the neutron star \citep{1996Natur.381..291F}. In particular,
the low observed projected semi-major axis ($a\sin{i}\sim2.6$\,light\,s) implies
a low-mass companion with a mass of $\le1M_\odot$ unless the system is viewed
almost precisely face-on \citep{1996Natur.381..291F}, which is unlikely for
a randomly oriented binary orbit (probability of less than 0.5\% for $i\le5.3^\circ$), and it also implies a companion mass of $\le1M_\odot$ (see, for instance, \citealt{1996Natur.381..291F}). The
evolutionary considerations also support a low-mass companion
\citep{1997ApJ...486..435R}, although the upper limit on the companion mass of
$\sim1.3M_\odot$ is also mostly imposed by the low probability of observing a binary
system face on in this case.

The low mass of the potential counterpart, and likely the location of the binary close
to the Galactic center, prevented any robust identification of an optical companion;
although, several candidates have been suggested \citep{1997ApJ...486.1013A,
2007MNRAS.380.1511G}. The most promising candidate was identified by
\citet{2007MNRAS.380.1511G} as a G/K~III star close to the Galactic center
based on the X-ray position of the source found using \emph{Chandra}
observations in quiescence \citep{2002ApJ...568L..93W} and VLT
spectro-photometry. With an estimated mass of $2-3M_\odot$, such a star would be
close to filling its Roche lobe, which would explain the observed
powerful X-ray outbursts. As discussed by \cite{2007MNRAS.380.1511G}, this
candidate appears to still be consistent with the evolutionary considerations
discussed by \cite{1997ApJ...486..435R}; although, a very low inclination is
still required to satisfy the observed mass function in this case. Therefore, another
nearby star was suggested as an alternative counterpart
\citep{2007MNRAS.380.1511G} at $\sim3.75$\,kpc. \cite{2007MNRAS.380.1511G} emphasized, however, that the companion cannot be robustly identified until an improved X-ray localization and deeper optical observations become available.

In the absence of a reliable optical association, most investigations of the source
properties have been conducted in the X-ray band, particularly during the two
outbursts observed in 1995 and 2014. An analysis of the 1995 outburst data made it possible to
determine the orbit of the system and to obtain an upper limit on the magnetic field of $\sim6\times10^{11}$\,G for the neutron star based on the nondetection of a propeller transition \citep{1996Natur.381..291F}. This conclusion appears to be consistent with evolutionary arguments \citep{1997ApJ...486..435R}, and it constitutes another argument for \src being a rather unusual source with the field being midway between the accreting pulsars and
millisecond accreting pulsars. These estimates were later confirmed by the claimed detection of the source transition
to the so-called propeller regime in the declining part of the 1995 outburst \citep{1975A&A....39..185I,1986ApJ...308..669S,1997ApJ...482L.163C}. The
``propeller'' effect, or the centrifugal inhibition of the accretion, sets in when
the magnetic field lines at the magnetosphere boundary move faster than the
Keplerian disk, which allows one to estimate the field from the observed luminosity
of the transition, defining the size of the magnetosphere. In particular,
\cite{1997ApJ...482L.163C} estimated the magnetic field at
$\sim2.4\times10^{11}$\,G. Finally, the magnetic field was also measured directly as $B\sim5\times10^{11}$\,G with the detection of a cyclotron line at $\sim4.3-4.7$\,keV during the same outburst
with \emph{BeppoSAX} \citep{2015MNRAS.452.2490D} and in 2014 with
\emph{XMM-Newton} \citep{2015MNRAS.449.4288D}. 
Besides that, extensive monitoring of the source with the Rossi X-ray Timing
explorer \citep[RXTE; ][]{1996SPIE.2808...59J}, the Compton Gamma Ray Observatory (CGRO), and other facilities has revealed a rich phenomenology for burst and persistent emission properties as well as their
dependence on luminosity
\citep{1998BAAS...30R.761F,1998AAS...192.6803S,1999ApJ...517..431W,2000ApJ...540.1062W, 2004AIPC..714..342F,2018MNRAS.481.2273C,2019MNRAS.482.1110J}, which is
still largely unexplored.

Another surprise came when \cite{2019A&A...626A.106M}
recently reported a discovery of a radiation-pressure dominated (RPD) accretion disk in this source based on the analysis of high-frequency variability as observed by RXTE. This made it possible to independently estimate the magnetosphere's size, which turned out to be significantly smaller than expected assuming a magnetic field estimate based on the observed cyclotron line energy. In particular, \cite{2019A&A...626A.106M} found that the effective magnetosphere size must be an order of magnitude smaller compared to the canonical Alfv\'en radius.
\cite{2019A&A...626A.106M} pointed to uncertainties in the interaction of the RPD disk with the magnetosphere or the presence of strong multi-pole components of the magnetic field as a reason for this discrepancy.

We note that comparatively small magnetospheric radii were also found by other authors.  Based on the analysis of the observed
spin-up rate dependence on luminosity during the 2014 outburst, \cite{2017MNRAS.469....2S} came to the same conclusion suggesting that $\xi\sim0.13-0.46,$ which corresponds to a distance of 5.1-3.4\,kpc, respectively, is required to reconcile the observed spin-up rate and cyclotron line energy. We note that the observed spin-up rate would imply even lower values for $\xi$ for the assumed distance of 8\,kpc. Under assumptions used by \cite{2017MNRAS.469....2S} to derive an analytic estimate for $\xi$, this would require, however, an unreasonably low disk viscosity, so the authors opted to put an upper limit on the distance instead. On the other hand, \cite{2017MNRAS.469....2S} ignored any possible braking torques reducing the expected spin-up rate, which could increase the magnetosphere's size. While accretion torque applied at the inner accretion disk is expected to be dominant at high accretion rates, we note that the deviation of the observed power law dependence of the spin-up rate on the flux from the expected 6/7 index \citep{1997ApJS..113..367B} indicates 
that a complete account of all the terms that drive the spin evolution of the source is clearly quite complex, and thus biases can be introduced when deriving the radius of the magnetosphere, for example.

Nevertheless,  evidence for a compact magnetosphere was also reported by \cite{2014ApJ...796L...9D} based on the 6.7\,keV iron line properties, and by  \cite{2015MNRAS.449.4288D} and \cite{2015ApJ...804...43Y} based on the properties of the thermal component likely coming from the inner edge of the accretion disk. In all cases, the effective magnetosphere size $R_{\rm m}$ was found to be only $\sim10-20$\% of the Alfv\'en radius for spherical  
\begin{equation} \label{ra}
 R_{\rm A}\simeq3.5\times10^8 \mu_{30}^{4/7} M^{1/7}R_6^{-2/7}L_{37}^{-2/7}\,{\rm cm}    
,\end{equation}
defining the radius where the ram pressure of the accretion flow and magnetic pressure are balanced. Here, $M$ and $R_6$ are the mass and radius of the compact object in units of solar mass and $10^6$\,cm, $\mu_{30}$ is the magnetic dipole moment in units of 10$^{30}$\,G\,cm$^3$, and $L_{37}$ is the accretion luminosity in units of 10$^{37}$\,erg\,s$^{-1}$. Generally, larger values are expected and a clear understanding of this discrepancy is, however, still lacking.  Independent estimates of the distance to the source and of the magnetic field are, therefore, desirable.

In this paper, we report on the deepest observations to date of the source in
quiescence with \emph{XMM-Newton} \citep{2001A&A...365L...1J}, which are aimed to provide a better localization of the source and to help us understand whether accretion continues in quiescence. As a result, we were able to confirm one of the previously identified optical counterpart candidates, which strongly suggests that \src is indeed located close to the Galactic center. We also discuss our results in the context of other investigations and available archival observations in quiescence. In particular, we find that the observed hard spectrum and strong variability strongly suggest that the accretion indeed continues down to luminosities of $\sim10^{34}$\,erg\,s$^{-1}$ (assuming distance of 8\,kpc), which also implies that the magnetosphere of the pulsar must be small compared to expectations based on the observed cyclotron line energy, and we discuss possible reasons for this.

\section{Observations and data analysis}
The source has mostly been observed during outbursts, however, some
observations in quiescence are also available. Two of those were used to
provide the most accurate X-ray localization to date, that is, with \emph{XMM-Newton}
\citep{2001A&A...365L...1J,2001A&A...365L..18S,2002A&A...386..531D} and \emph{Chandra} \citep{2000SPIE.4012....2W,2003SPIE.4851...28G,2002ApJ...568L..93W}.
Surprisingly, these are not the only detections of the source in quiescence, which was routinely serendipitously observed by \emph{XMM-Newton} as part of the Galactic center monitoring program. In all but one case, the source was significantly detected; the exception being observation 0112971201 where only PN was operating and the orbital background was quite high.

Here we report on 62\,ks observation of the source in quiescence with
\emph{XMM-Newton} (the deepest to date), which was performed on Apr 4, 2019
(obsid. 0821120101). The observation was conducted in full-frame mode for EPIC
PN \citep{2001A&A...365L..18S} and small-window mode for MOS \citep{2001A&A...365L..27T} to avoid a possible pileup. The data reduction
was done using the XMM SAS version 18.0 and the current set of calibration files as
of Feb 2020 following recommended procedures. Besides that, we use the fluxes
reported in the 4XMM-DR9 catalog \citep{2009A&A...493..339W,2016A&A...590A...1R} for
other observations. 

\subsection{Updated X-ray position and the optical counterpart}
\label{sec:oc}
The location of \src in the crowded region close to the Galactic center complicates
the identification of the optical counterpart. The most accurate X-ray localization
of the source so far has been reported by \cite{2002ApJ...568L..93W} based on the
\emph{Chandra} observation in quiescence, which was used by
\cite{2007MNRAS.380.1511G} to identify two potential counterparts within
$\sim0.8^{\prime\prime}$ X-ray error circle. An improved X-ray localization of
the source is, therefore, necessary to discriminate between the two.
The \emph{Chandra} pointing uncertainty dominates the accuracy of the
localization and can only be
improved through field rectification by matching the position of
other detected X-ray sources with those of their optical counterparts.
This procedure can actually be easier for \emph{XMM-Newton} with its larger field of view, and it is routinely used, for instance, in the \emph{XMM-Newton} serendipitous source
catalog (SSC) \citep{2009A&A...493..339W,2016A&A...590A...1R}.
The latest release of this catalog, 4XMM-DR9, in fact already contains seven
measurements of the X-ray position for \src with an estimated accuracy, including
the remaining systematics, that ranges from $\sim0.65^{\prime\prime}$ to
$\sim1.5^{\prime\prime}$, that is to say it is comparable with that of \emph{Chandra}. The
latest and deepest observation analyzed in this work is, however, not included
in this catalog. We followed, therefore, a similar pointing rectification procedure as implemented by \cite{2009A&A...493..339W} to also obtain the position of
the source for this observation.

In particular, after running the screening for periods of high in-orbit
background with tasks \texttt{pn-filter/mos-filter}, which reduced
the effective exposure to $\sim28$\,ks for EPIC~PN and to $\sim50$\,ks for MOS, an
image of the field was extracted. We then ran the source detection chain
\texttt{edetect\_chain} in the 1-7.2\,keV band (the lower energy band is
strongly affected by absorption) for all three cameras, which revealed 59 X-ray sources with potential optical counterparts within
10$^{\prime\prime}$ from the X-ray position in the Gaia DR2 catalog
\citep{2018A&A...616A...1G}. We then matched the X-ray and optical positions,
obtaining the corrected X-ray positions for all sources in the field using
the SAS task \texttt{eposcorr}. The residual systematics for matched sources
was estimated at 0.35$^{\prime\prime}$ (i.e., comparable to that reported for
4XMM-DR9), using the task \texttt{evalcorr}, and added in quadrature to the
statistical uncertainty. Thus for \src, we find RA=17:44:33.08 and
Dec=-28:44:26.93 with the uncertainty of 0.38$^{\prime\prime}$. We note that our estimate is
still dominated by systematics, but it has substantially improved compared to \emph{Chandra}.

\begin{figure}[ht]
        \centering
                \includegraphics[width=\columnwidth]{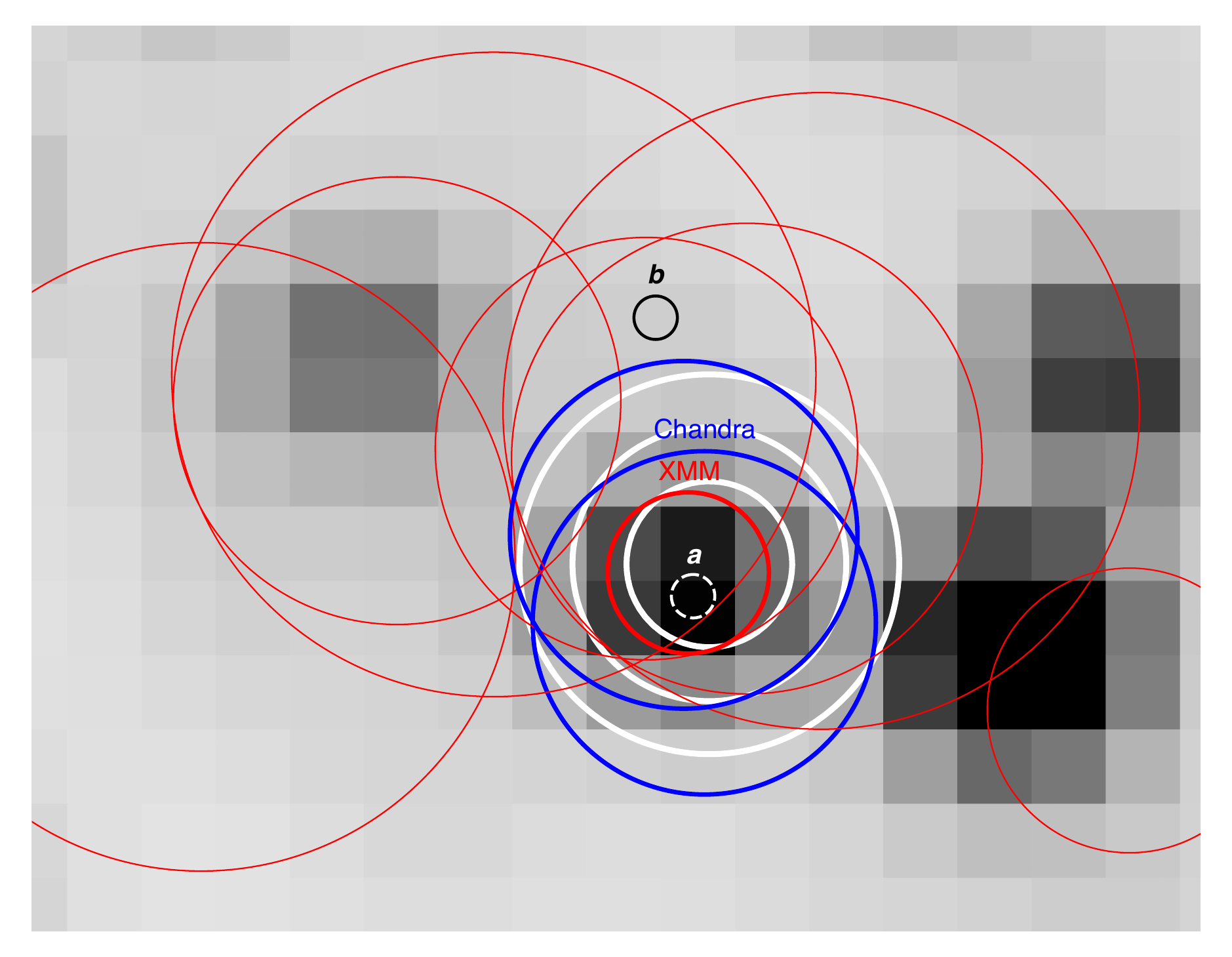}
        \caption{Deep stack $K_s$ band image from the VVV survey \citep{2010NewA...15..433M}. Location of the two counterparts (labeled \emph{a} and \emph{b} in the picture with respective uncertainties indicated by circle radii) suggested by \cite{2007MNRAS.380.1511G} with X-ray positions of \src, reported by \cite{2002ApJ...568L..93W}, and in the CSC 2.0 catalog (blue), the 4XMM-DR9 catalog (thin red lines), and as obtained in this work (thick red line). The white  contours indicate $1-3\sigma$ confidence levels  from the joint fit of all aforementioned X-ray positions obtained as described in the text. }
        \label{fig:loc}
\end{figure}

To further improve the accuracy, in addition to our new estimate, we also considered all previously reported \textit{XMM-Newton} and \textit{Chandra} positions from catalogs. For each measurement a $\chi^2$ statistics was
calculated as $\sum_{i=1}^{9}{(\Delta_i/\sigma_i)^2}$, where $\Delta_i$ is
the distance between a given point and an individual position estimate, and $\sigma_i$
is the reported uncertainty for the corresponding measurement, which includes the estimated systematics. This fit yielded the
final position of RA=17:44:33.08 Dec=-28:44:27.03, corresponding to a
$\chi^2$ minimum value of 16.7 (for 7 degrees of freedom). The final positional uncertainty can then be estimated by comparing this value with the statistics at an arbitrary distance, that is, projection, by finding distances corresponding to change in the statistic value of $\Delta\chi^2$=2.3, 6.18, and 11.83 for $1-3\sigma$ confidence levels, respectively.
The resulting contours are shown in Fig.~\ref{fig:loc} and can be approximated with circles with radii of $\sim0.37^{\prime\prime}$ ($1\sigma$ confidence) to $\sim0.89^{\prime\prime}$
($3\sigma$ confidence).
We emphasize that this estimate already accounts for systematic uncertainties of individual positional measurements.
One might argue that the final fit implies relatively large reduced
statistics, which is, however, driven by a single, strongly deviating position
reported in the 4XMM-DR9 catalog (for obsid. 0506291201), which is inconsistent with both
positions found for \emph{Chandra} and \emph{XMM-Newton}  in this work. The source is
located, however, close to the edge of the field of view in this observation,
so the accuracy of the position may be compromised and thus it is likely an
outlier. We still include it in the fit for completeness, but emphasize that
exclusion of this observation would move the location of the source closer to what was
found by us and \emph{Chandra}; additionally, at the same time, it would slightly reduce the uncertainty. Even if this position is included, however, the alternative optical
companion (companion
\emph{b}) suggested by \cite{2007MNRAS.380.1511G} is still ruled out at more than $3\sigma$ confidence. The comparatively massive G/K~III star is thus the preferred counterpart. As discussed by \cite{2007MNRAS.380.1511G}, this implies that the system must be viewed essentially face on, and the distance to the optical counterpart is consistent with a distance to the Galactic center in this case. 

\subsection{Timing analysis}
 After screening  data for periods of high in-orbit background as described
above, the scientific data products (i.e., source event lists, light curves, and
spectra) for \src were extracted individually for each of the three cameras from the region centered on the source. The
extraction region size was optimized using the \texttt{eregionanalyse} task
and was found to be 26$^{\prime\prime}$ and 28$^{\prime\prime}$ for PN and MOS,
respectively. The background spectrum was extracted from nearby regions on the
same chip. For MOS, we used an annulus with an inner radius of 45$^{\prime\prime}$ and an
outer radius that was large enough to include all counts detected in the readout window. 
For PN, we used a circle with a radius of 45$^{\prime\prime}$ located on the
same chip as the source and at the same distance along the Y instrumental
coordinate as recommended in the documentation of the instruments. The background-subtracted source light curves with a time resolution of 1\,ks in the 0.5-10\,keV band were then obtained for all three cameras using the task \texttt{epiclccorr,} which performs background subtraction and corrects for difference in effective areas of the source and background and other instrumental effects. Finally, to improve the counting statistics, we co-added the resulting light curves using the \texttt{lcmath} task.

\begin{figure}[!t]
        \centering
        \includegraphics[width=\columnwidth]{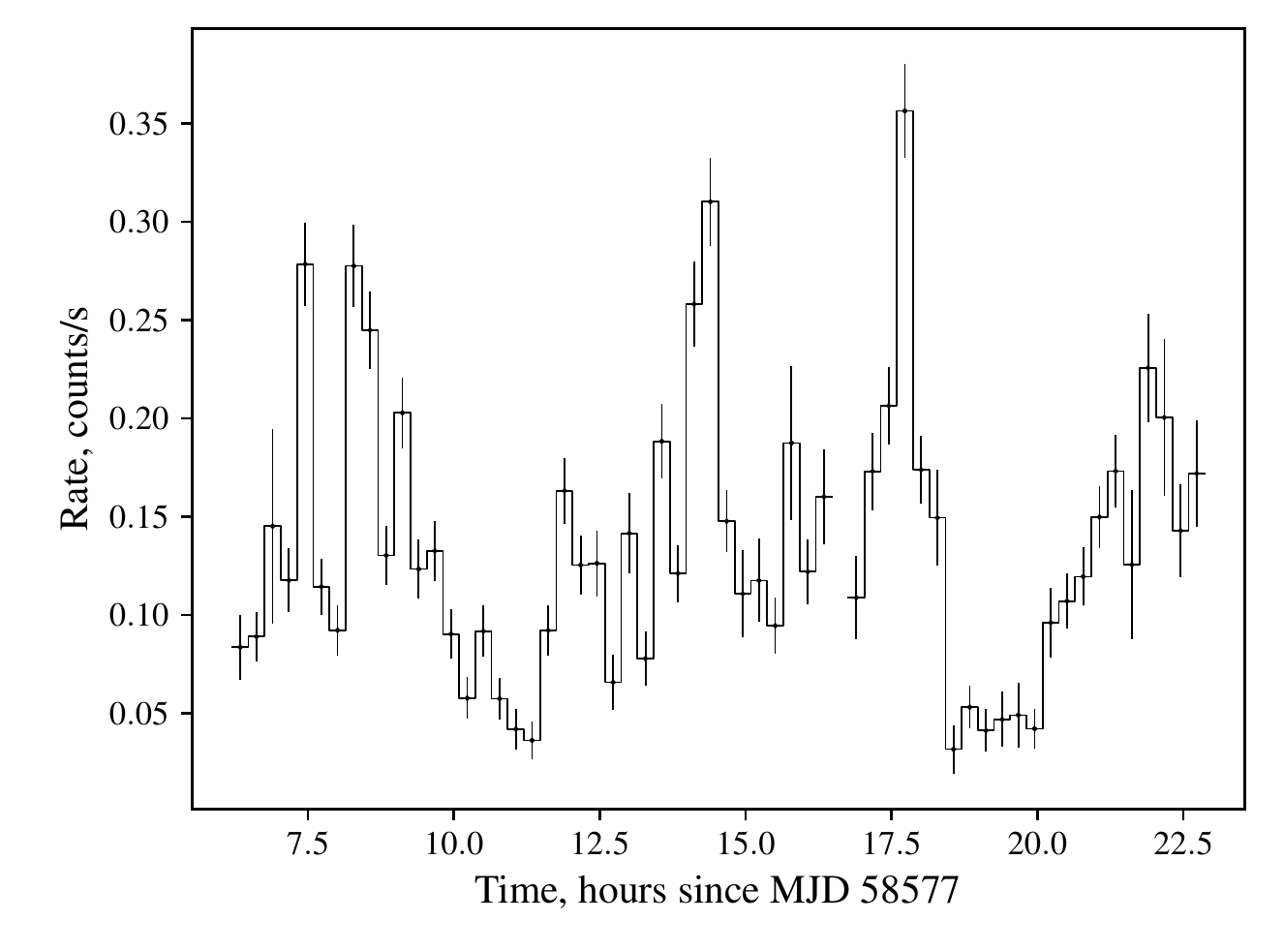}
        \caption{Light curve of \src with time resolution of 1\,ks as observed by \textit{XMM-Newton} in the 0.5-10\,keV energy band (all detectors combined) reveals strong flux variability.}
        \label{fig:xmmlc}
\end{figure}

As is evident from Fig.~\ref{fig:xmmlc}, the resulting light curve exhibits significant variability with the flux changing by almost an order of magnitude during the observation. Such variability can only be expected from an accreting neutron star and thus strongly suggests that \src indeed continues to accrete during the \textit{XMM-Newton} observation.

\begin{figure}[!t]
        \centering
                \includegraphics[width=\columnwidth]{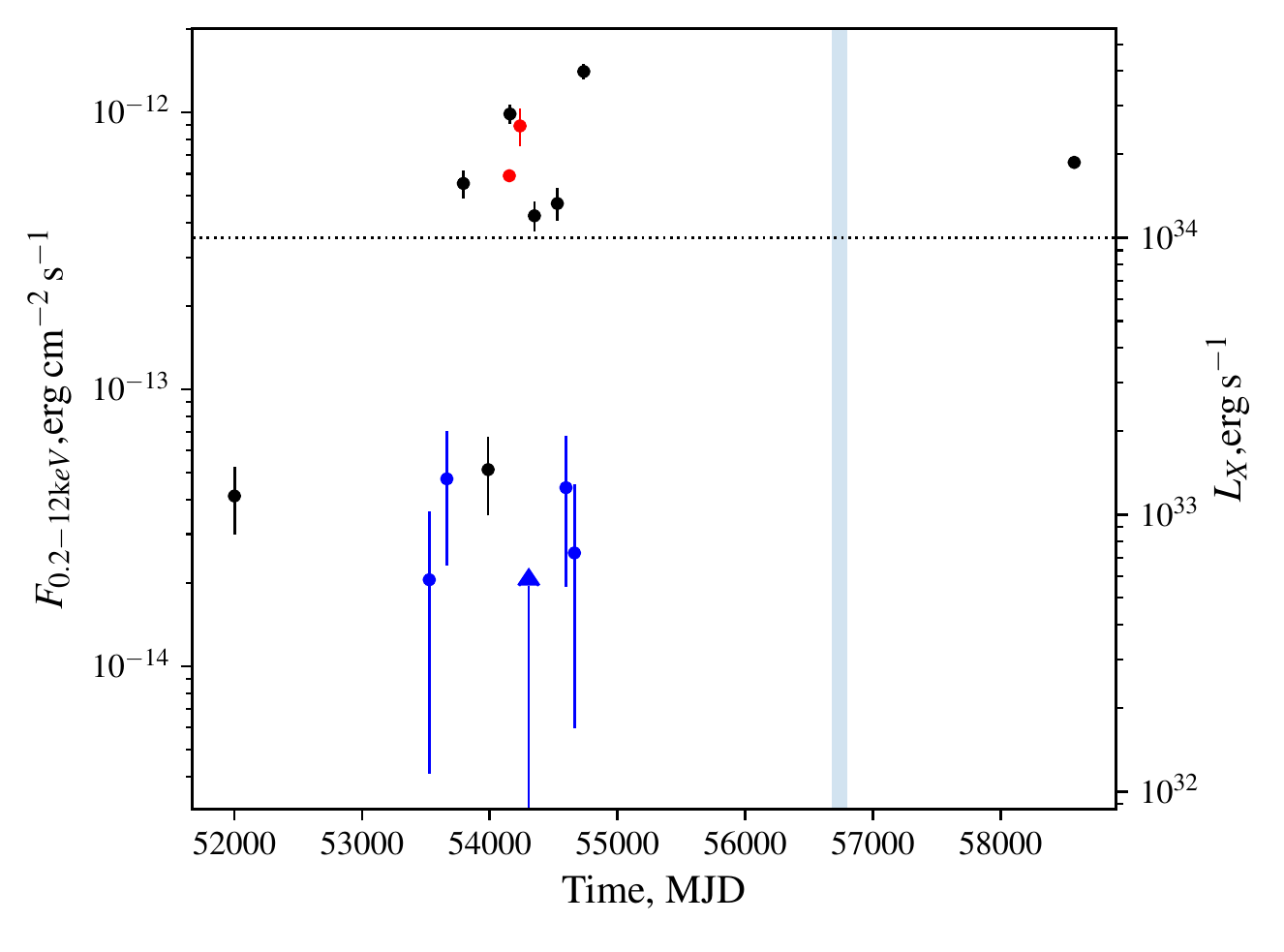}
        \caption{The historical light curve of \src as observed by \emph{XMM-Newton} (black points) and \emph{Chandra} (blue points for HRC and red for ACIS-S cameras). Here, the bolometric luminosity estimated, assuming a distance to the source of 8\,kpc and a bolometric correction factor of 3.7, is also indicated. The shaded area indicates the 2014 outburst of the source.
        The dotted line indicates the propeller luminosity estimated in this work.}
        \label{fig:all_lc}
\end{figure}

It is also interesting to re-visit other observations of the
source in quiescence to investigate variability on longer timescales. Being
close to the Galactic center, \src has been observed several
times with \emph{XMM-Newton} and \emph{Chandra} and detected in the majority
of observations. To investigate the source variability, we used, therefore, broadband
fluxes reported in \emph{XMM} SSC
\citep{2009A&A...493..339W,2016A&A...590A...1R} and in the \emph{Chandra} source catalog
\citep{Evans_2010}. Considering that \emph{Chandra} fluxes are reported in the
0.5-7\,keV band and \emph{XMM-Newton} fluxes are in 0.2-12\,keV, for comparison purposes we applied
a correction factor of 1.53 in the former case as estimated from the best-fit
spectrum for the accretion-heated model. The results are presented in Fig.~\ref{fig:all_lc}. It is interesting
to note that the distribution of observed fluxes appears to be bimodal, clustering
around $\sim10^{-12}$\,erg\,cm$^{-2}$\,s$^{-1}$ and
$\sim10^{-14}$\,erg\,cm$^{-2}$\,s$^{-1}$, respectively, that is, the distribution varies by more than
an order of magnitude on a timescale of several hundred days. Again, such variability can only be explained by accretion, so we conclude that also historical observations of the source in quiescence strongly imply that \src continues to accrete during the observations belonging to the group with a higher flux, which includes the observation reported in this work. It is also interesting to note that during the outbursts, persistent emission of \src appears to be much less variable than we observe in quiescence, which could point to unstable accretion, as is expected around the transition to the propeller regime.

In this context, it is interesting to note that the observed historic flux distribution of \src resembles the one reported by
\cite{2016MNRAS.457.1101T} for the pulsating ultra-luminous X-ray source M82
X$-$2, 
where it was used to argue for the onset of the centrifugal inhibition of the
accretion, that is, the transition to a propeller regime
\citep{1975A&A....39..185I}. Furthermore, the observed luminosity gap of $\sim100$ matches the expected change in luminosity upon the transition to the propeller for the observed period of the source \citep{1996ApJ...457L..31C}. We suggest, therefore, that the two peaks in the observed flux distribution correspond to the accretion and propeller states, respectively. 


\begin{figure}[!t]
        \centering
                \includegraphics[width=\columnwidth]{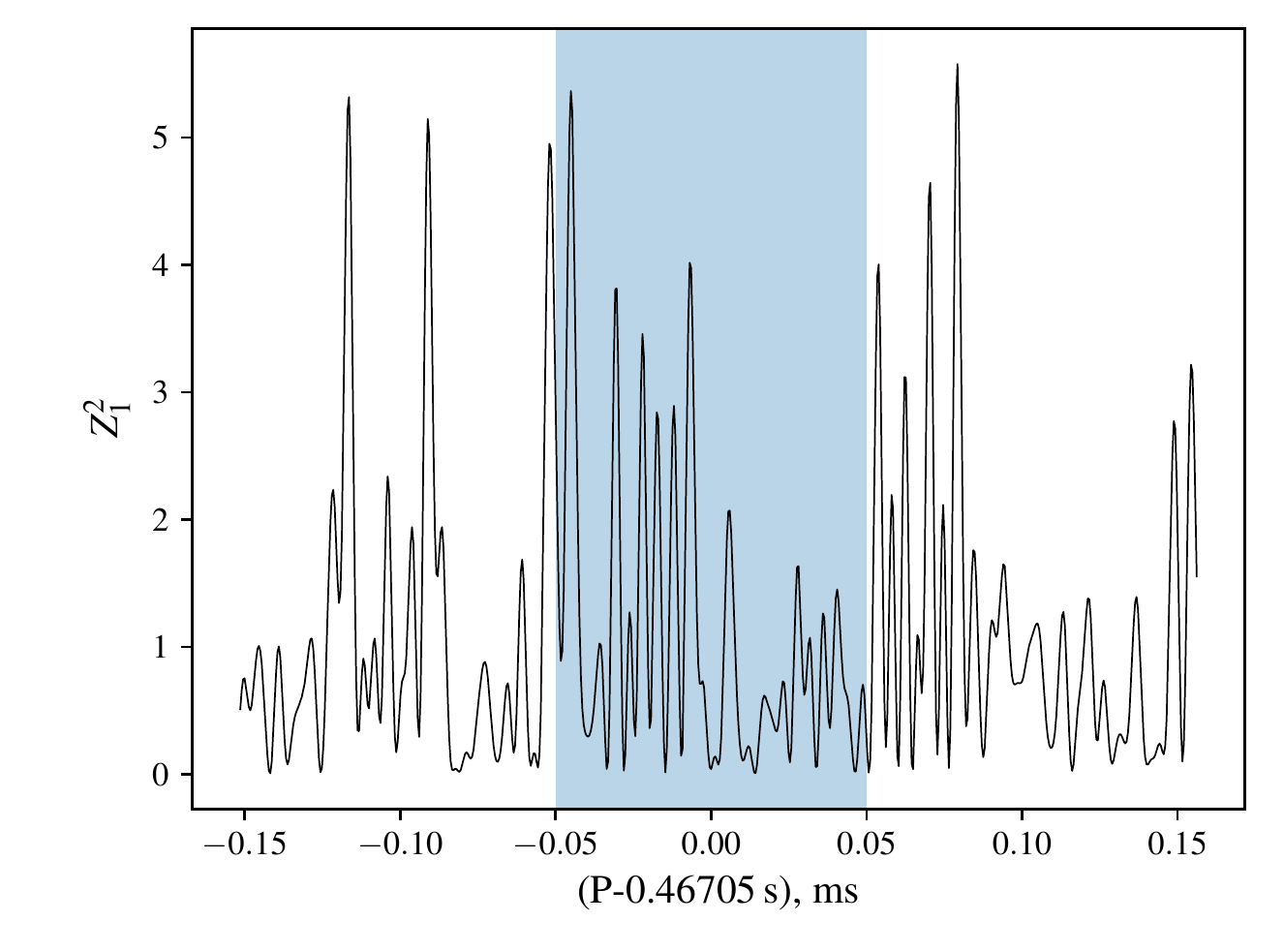}
        \caption{Periodogram for EPIC~PN data in 2-7\,keV energy range. The shaded area represents the historically reported period range.}
        \label{fig:periodogram}
\end{figure}

The detection of the pulsations with an amplitude and pulse profile shape that are similar to what was observed in the outburst would also support an accretion-powered origin of the observed emission. We conducted, therefore,
a periodicity search to detect this type of signal. The analysis was conducted using EPIC PN data only because the time resolution of
the MOS cameras (0.3\,s in small window mode) is insufficient to detect pulsations
with an $\sim0.467$\,s period expected from the source.
To search for pulsations, we used source photons extracted from the same region, rather than light curves, to prevent sensitivity loss due to the time binning. The energy range for the pulsations search was limited to 2\,keV to 7\,keV, where most
source photons are detected. To increase the number of photons and avoid gaps in
the light curve, we also ignored the high in-orbit background periods mentioned
above for this analysis. We verified, however, that this does not affect the
results by repeating the analysis for screened data. As a result, 3324 photons were
selected. Based on the spectral analysis of the results described below, $\sim500$ of these must be background photons. The photon arrival times were then corrected to the solar
barycenter, using the \texttt{barycent} task, and for binary motion, using
the ephemerides reported by \cite{2017MNRAS.469....2S}. The pulsation search was
conducted using the $Z^2$ statistics \citep{1989A&A...221..180D} for periods in
the range between 0.466\,s and 0.468\,s, which is significantly wider than periods
historically reported from the source. No significant peaks were found, as
is illustrated in Fig.~\ref{fig:periodogram}. 

To estimate the sensitivity of our
observation to pulsations, we used the method proposed by
\cite{1994MNRAS.268..709B} as described in \cite{2015A&A...579A..22D}, which
permitted us to put an upper limit on the detectable pulsation amplitude (defined as $(\max-\min)/(\max+\min)$ of the count rate over the folded light curve) of $\sim20$\%. We note that the pulsation amplitudes reported for the outburst range between 5 and 20\% in the 2-10\,keV energy range, and they increase with energy \citep{2015MNRAS.452.2490D}, that is, they are consistent with a deduced upper limit. Furthermore, the maximal pulsed fraction expected from a rotating neutron star with two thermally-emitting hotspots \citep{2002ApJ...566L..85B} is comparable to the limit obtained above.
We conclude, therefore, that the lack of a clearly detected pulse signal is unsurprising, given the available statistics for this observation.

\subsection{Spectral analysis} 
\begin{figure}[!t]
        \centering
                \includegraphics[width=\columnwidth]{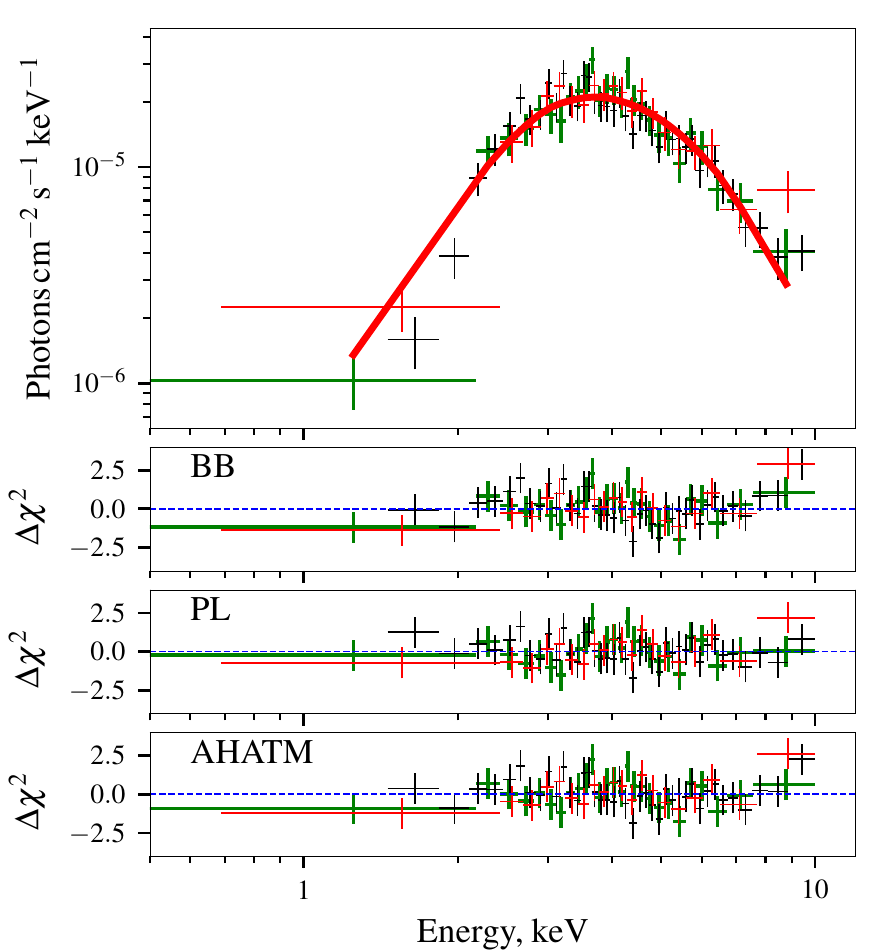}
        \caption{Best-fit unfolded spectrum (with absorbed blackbody model, top), and corresponding residuals for fits with absorbed blackbody (\texttt{BB}), power law (\texttt{PL}), and accretion heated atmosphere (\texttt{AHATM}) models.}
        \label{fig:spe}
\end{figure}

Our conclusion that the source continues to accrete is also supported by the spectral analysis. As described above, the spectra of the source were extracted individually for each of the three cameras.
The source count rate
was found to be 0.02 and 0.06 for MOS and PN, respectively, which constitutes
$\sim80$\% of the total count rate. This implies that the observation is likely not
affected by pile-up. Nevertheless, we confirmed this using the task
\texttt{epatplot}. The background, however, dominates the source count rate
below $\sim1$\,keV, especially for PN, which is due to the strong absorption in
the direction of the source and the residual contribution of the soft proton-induced
background. We restricted, therefore, the analysis of the energy band to above
0.5\,keV for MOS and above 0.9\,keV for PN. The unfolded spectrum of the source
and best-fit residuals for several models is presented in Fig.~\ref{fig:spe}.
\begin{table}
        \begin{center}
        \begin{tabular}{llll}
                Parameter/model & \texttt{BB} & \texttt{PL} & \texttt{AHATM} \\
                \hline
                $N_{\rm H}, 10^{22}$\,cm$^{-2}$ & 7.1(5) & 13.8(5) & 8.6(4)\\
                $kT$\,keV/$\Gamma$/$\ell_{\rm a}$ & 1.25(4) & 2.6(1) & 0.13(1)\\
                $A$/10$^{-3}$ & 34(5) &  1.4(4) &   \\
                $C_{\rm M1}$& 1.05(5) & 1.05(5) & 1.05(5)\\
                $C_{\rm M2}$& 1.03(5) & 1.03(5) & 1.03(5)\\
                $F_{\rm x,obs}$,10$^{-13}$\,erg\,cm$^{-2}$\,s$^{-1}$ & 6.4(2) & 6.8(2) & 6.6(2) \\
                $F_{\rm x,src}$,10$^{-13}$\,erg\,cm$^{-2}$\,s$^{-1}$ & 9.9(4) & 48(8) & 11.8(4)\\
                $\chi^2$/dof & 86.36/86 & 63.72/86 & 73.51/86\\
                
        \end{tabular}
        \end{center}
        \caption{Best-fit parameters for \src spectrum as observed with \emph{XMM-Newton} (obsid. 0821120101) and described with an absorbed blackbody (\texttt{BB}), power law (\texttt{PL}), and accretion-heated atmosphere (\texttt{ATM}) models. The fit statistics and estimated source flux in the 0.5-10\,keV energy band (observed and unabsorbed) are also listed for each model.}
        \label{tab:speres}
\end{table}

Considering the available counting statistics and the limited energy band covered by
\emph{XMM-Newton}, several models can be used to describe the observed spectrum. In
principle, both the absorbed blackbody (\texttt{tbabs}$\times$\texttt{bbodyrad}
in {\sc xspec}) and an absorbed power law
(\texttt{tbabs}$\times$\texttt{powerlaw} in {\sc xspec}) provide a reasonable
description of the spectrum. From a purely statistical point of view the power law fit provides the best fit as summarized in Table~\ref{tab:speres} and
Fig.~\ref{fig:spe} (all uncertainties are quoted at $1\sigma$ confidence unless
stated otherwise). 

No evidence for the presence of an absorption feature is evident
from the residuals. That is actually not surprising since the feature was also not
detected in the fainter of the three \emph{BeppoSAX} observations
\citep{2015MNRAS.452.2490D} due to the low counting statistics. When a feature is formally included in the fit as a multiplicative Gaussian in absorption (\texttt{GABS} in {\sc xspec}) with the energy and width fixed to that reported by \cite{2015MNRAS.452.2490D}, the resulting line depth is consistent with zero. On the other hand, the $3\sigma$ upper limit on depth of $0.28$ is above the value reported in the same paper (0.12), so we conclude that our \emph{XMM-Newton} observation is not sensitive to the potential presence of a line with reported parameters and thus the presence of the feature cannot be ruled out. 

We note that  the black-body fit exhibits a systematic structure in residuals, and, what is more important, it yields a rather high temperature of $\sim1.2-1.3$\,keV compared to $\sim0.5$\,keV, which is typically reported for cooling neutron stars. That is actually not surprising considering the observed variability of the source, which strongly suggests that \src is in fact accreting and is in line with the fact that the power law fit
is clearly superior from  a statistical point of view.

On the other hand, the resulting value of the absorption column
($\sim14\times10^{22}{\,\rm cm}^{-2}$) is substantially higher compared to
values reported by \cite{2015ApJ...804...43Y} and \cite{2015MNRAS.449.4288D}
for the 2014 outburst based on the broadband spectral analysis ($8.8(3)\times10^{22}$\,cm$^{-2}$ and $6.1(3)\times10^{22}$\,cm$^{-2}$
respectively, the same absorption model was used).\ This might indicate that there are either changes in intrinsic absorption within the system or problems with the simple power law model.
Along with a rather steep resulting power law index, the later option seems to be more realistic and points to a more complex spectral shape.
\begin{figure}[ht]
        \centering
                \includegraphics[width=0.95\columnwidth]{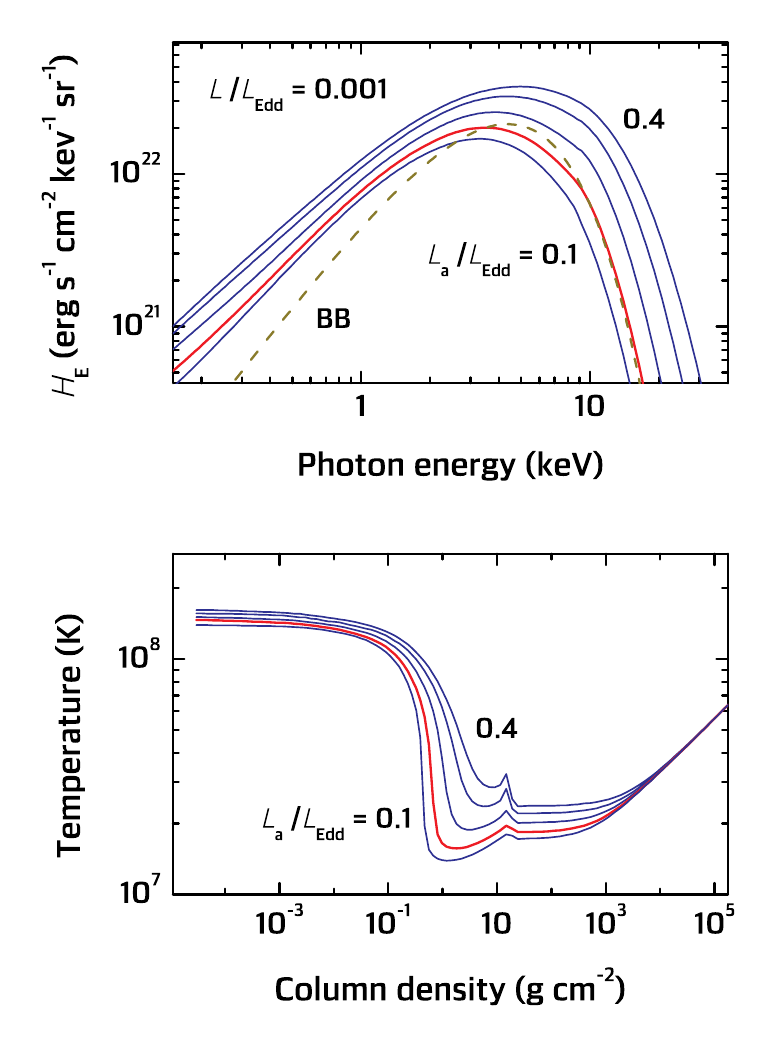}
        \caption{Spectra of the computed accretion heated atmospheres (top panel) and the
        corresponding temperature structures. The grid models with the relative accretion luminosities $L_{\rm a} / L_{\rm Edd} =$\,0.1,0.2,0.3, and 0.4 are
        shown with blue curves. The best fit model spectra and temperature structure are shown with red curves. The best fit diluted blackbody spectrum in the NS surface frame 
        ($T'_{\rm BB}$=1.545\,keV) is also shown
        with a dashed curve.
        }
        \label{fig:aheat}
\end{figure}

We attempted, therefore, to fit the spectrum of the source with a physical
model aimed to describe the X-ray spectrum of a neutron star atmosphere heated
by an accretion flow. 
The detailed description of the model can be found in the paper by \citet{2018A&A...619A.114S}. The code described in the paper allowed us to compute the NS model atmosphere that is heated by the accretion flow by considering the accreting ions as independent fast-moving particles penetrating the atmosphere. The input parameters of the model are the NS mass $M$ and
radius $R$, the relative intrinsic luminosity $\ell=L/L_{\rm Edd}$, the relative luminosity due to
accretion heating $\ell_{\rm a}=L_{\rm a} / L_{\rm Edd}$, and the chemical composition of the atmosphere.
The input parameters of the accretion flow are the relative velocity of the ions 
$\eta={\rm v}/{\rm v}_{\rm ff}$, where ${\rm v}_{\rm ff}$ is the free-fall velocity at the NS surface,
the relative temperature of the ions $\chi=T/T_{\rm vir}$, where $T_{\rm vir}$ is the virial
temperature of the ions at the NS surface, and $\Psi$ is the penetration angle of the ions
into the atmosphere. 

It is important to note that the model does not take the magnetic field into account.
We emphasize that this approach leads to potentially significant uncertainties
because the cyclotron energy in the spectrum at the high luminosity state at $\sim4-5$\,keV is in the
range where most of the flux emerges, so our results should only be considered as a first approximation. 
If the magnetic field is important, \src could be  
similar to other X-ray pulsars in low states, such as X~Per \citep{Doroshenko.etal:12},
A 0535+262 \citep{Tsygankov.etal:19b}, and GX~304$-$1 \citep{Tsygankov.etal:19}. 
These X-ray pulsars exhibit a two component spectra, where the high energy component is likely associated with cyclotron emission that is broadened with thermal Compton scattering in an overheated atmosphere as it also happens in our nonmagnetic models (see Fig.\,\ref{fig:aheat}).  For \src, the cyclotron emission is expected to fall into the soft X-ray band and may actually represent the main source of seed photons, which would significantly modify the emerging spectrum in analogy to the high energy component in the aforementioned highly magnetized pulsars. Unfortunately, our model is unable to account for this effect as of yet, so we have to resort to a low magnetic-field approximation.
On the other hand, 
as discussed below, the magnetic field of the neutron star could be substantially lower, about $\sim 10^{10}$\,G, and in which case
our results should be reasonably accurate.

We chose the following values of the parameters: $M = 1.4 M_\odot$, $R = 12$\,km,
and the normal in-falling  of the cold free-falling accreting ions,
$\Psi = 0\degr$, $\eta = 1$, and
$\chi = 0.001$. The chemical composition of the atmosphere as well as the accreting plasma 
were taken to be solar.  We then computed a grid of the models with the accreting heating luminosities
$\ell_{\rm a} =$\,0.1, 0.2, 0.3, and 0.4, which, along with normalization, are thus the only free parameters that varied during the fit procedure. The intrinsic luminosity of the model atmosphere was
taken to be $\ell = 0.001$. The emergent spectra and the temperature structures of the
models are shown in Fig.\,\ref{fig:aheat}. The computed spectra were transferred to the
distant observer frame, $E'=E(1+z)$ and $F_{E'} = 4\pi H_{E}/(1+z)^3$, where 
$1+z$\,=\,$(1-2GM/Rc^2)^{-1/2}$\,= 1.236 for the chosen NS model. 
The model grid was implemented as an {\sc xspec} table model and used to fit the
spectrum. The accretion heated atmosphere model provides a better description
of the \emph{XMM-Newton} spectrum compared to a pure blackbody as well as a lower absorption
compared to an absorbed power-law model, which still provides, however, a statistically superior fit. We emphasize that regardless of the model used, the observed spectrum is more compatible with what is expected from an accreting rather than thermally emitting neutron star.

However, using a physical model allows one to make some useful estimates and check for the self-consistency of the model.
The best-fit local accretion rate $\ell_{\rm a} =$\,0.134
and this computed model are also shown in Fig.\,\ref{fig:aheat}. This relative 
accretion rate corresponds to the effective temperature $kT_{\rm eff}$\,=1.164\,keV
or the observed one of 0.942\,keV. The diluted 
blackbody spectrum with the back redshifted observed temperature 
$T'_{\rm BB} = 1.25$\,keV\,$(1+z)$\,= 1.545\,keV is shown as well. The dilution factor,
0.32, is close to the expected value from estimated color correction $f_c = 1.25/0.942$\,=1.33,
$w \approx f_c^{-4}$. The derived normalization, $1.38 \times 10^{-36}$, makes it possible to find the 
radius of the radiating spot, $R_{\rm sp} \approx 4.1 \times 10^4$\,cm, assuming the distance to the source equals 8\,kpc and that the emission region is the flat disk inclined with the 
angle $60\degr$ to the line of sight. 
It is quite likely that the magnetospheric radius is close to the corotation one at this low luminosity, 
$R_{\rm m} \approx R_{\rm c} \approx 10^8$\,cm. In accordance with \citet{2015MNRAS.447.1847M}, the single hot spot area is $A = 2\pi R^3/R_{\rm m} 
\times (H/R_{\rm m})$. Comparing it with the observed hot spot area $\pi R_{\rm sp}^2
\approx 5.3 \times 10^9$\,cm$^2$, we can estimate the relative disk half-thickness at
the magnetospheric radius $H/R_{\rm m} \approx$\,0.05. The expected value at
the corotation radius is about 0.03 \citep[see, e.g.,][]{2007ARep...51..549S}, assuming the standard accretion $\alpha$-disk model \citep{1973A&A....24..337S}, so the hotspot size estimated above is in line with expectations. We also note that the effective temperature at the corotation radius is about 60\,kK for a given $\dot M$. Therefore, the disk is not in a cold state here, so a transition to the propeller regime is indeed expected as the accretion rate further decreases \citep{2017A&A...608A..17T}.
Thus, we conclude that the heated atmosphere model provides a physically
sound description of the observations and supports accretion scenario. We note, however, that broadband
observations of the source in quiescence would be more constraining and are
important to verify this conclusion and the model itself.

\begin{figure}[ht]
        \centering
                \includegraphics[width=\columnwidth]{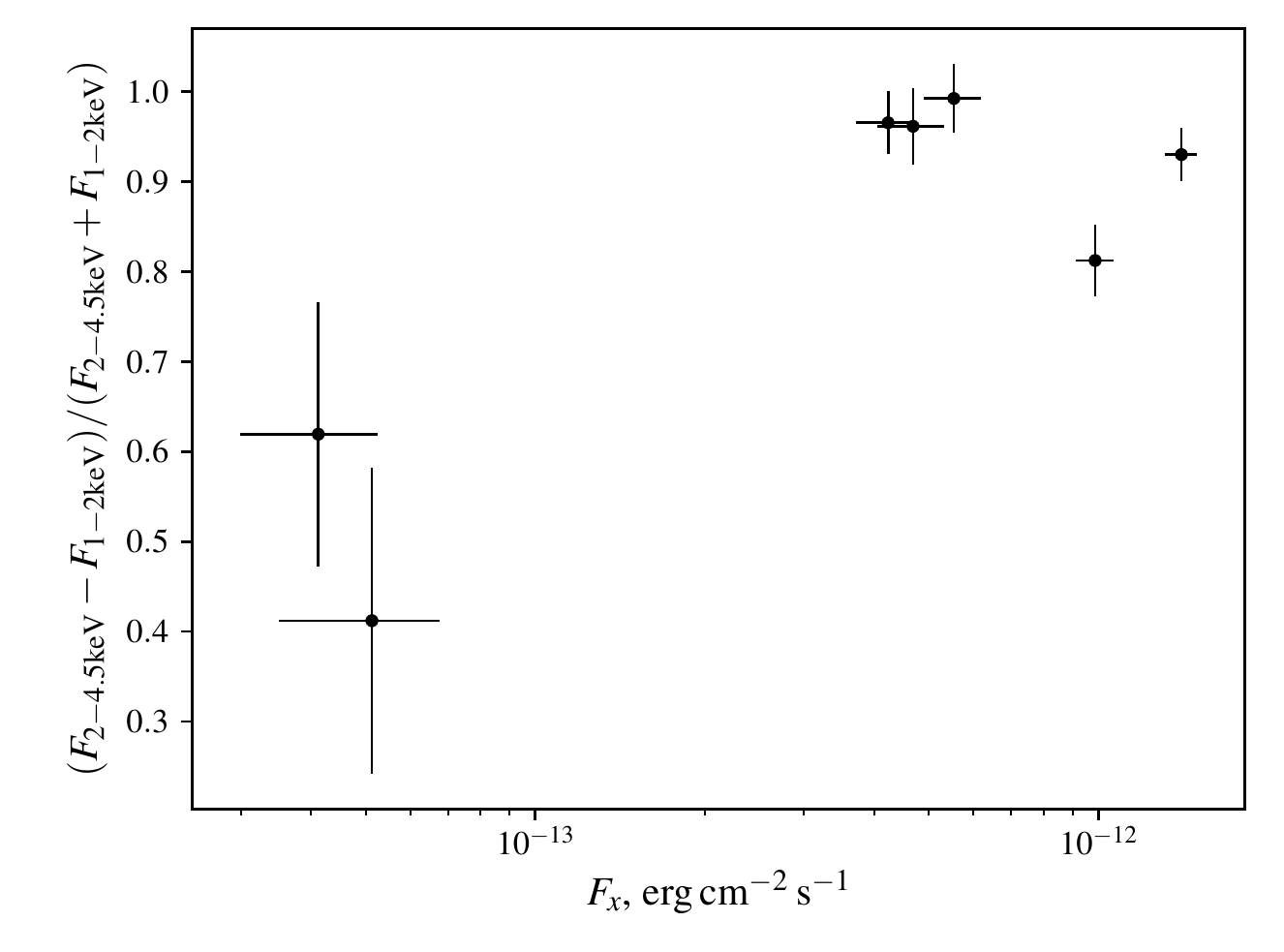}
        \caption{Evolution of spectral hardness reported in the \textit{XMM SSC} catalog for \src (\texttt{HR3} field) as a function of flux.}
        \label{fig:hr}
\end{figure}

Finally, it is also interesting to note that the source appears to soften at the lowest fluxes, as can be illustrated in Fig.~\ref{fig:hr}.
The spectra
extracted from the source position for the two observations with the lowest flux
(obsid. 0112971901 and 0302884201) contain just over $\sim50$\, source
counts in total  and thus they are hardly usable for a meaningful spectral analysis. Nevertheless, when described
with an absorbed blackbody model (with a fixed absorption column), the best-fit
temperature decreases to $\sim0.8(1)$, which is close to values reported by
\cite{2017MNRAS.470..126T} for propelling sources. 
We emphasize, however, that in this case
much deeper observations would also be required for a detailed analysis and we prefer to regard hardness as the main indicator of spectral softening.

\section{Discussion} The transition of \src to the propeller regime has, in fact, been already claimed
by several authors \citep{1997ApJ...482L.163C,2018MNRAS.477L.106C}, albeit for
significantly higher luminosities of $\sim0.8-2.5\times10^{37}$\,erg\,s$^{-1}$ \citep{1997ApJ...482L.163C,2018A&A...610A..46C}. On the other hand, the lowest flux for the higher flux group of quiescent detections of \src
with \emph{Chandra} and \emph{XMM-Newton} correspond to a luminosity of $\sim10^{34}$\,erg\,s$^{-1}$. Here, we applied a bolometric correction
factor of 3.7 to the observed flux in the 0.2-12\,keV band as estimated based on the spectral
parameters reported by \cite{2015MNRAS.452.2490D} during the outburst. We emphasize again that the source is still detected even significantly below this level, and such variability is inconceivable for the thermal emission from a thermally-emitting neutron star expected if the propeller transition would already take place at $10^{37}$\,erg\,s$^{-1}$.
We conclude, therefore, that the true propeller luminosity, that is, the accretion luminosity corresponding to the onset of the propeller effect for \src is $\sim10^{34}$\,erg\,s$^{-1}$, which is three orders of magnitude lower than previously suggested.

The new estimate is, however, significantly lower than theoretical expectations assuming the field derived from cyclotron line energy. Indeed, the propeller
luminosity can be estimated by equating the magnetospheric and corotation radii \citep{2018A&A...610A..46C}:
\begin{equation}
    L_{\rm prop} \simeq 1.97\times10^{38}\xi^{7/2}\mu_{30}^2 P^{-7/3}M^{-2/3}R_6^{-1}\,{\rm erg\,s}^{-1},
    \label{eq:lprop}
\end{equation}
where $\xi$ is the ratio of the magnetospheric radius to the Alf\'ven radius  and $P$ is the spin period of a neutron star in seconds. This relation was observationally calibrated by \citet{2018A&A...610A..46C}, using multiple types of magnetized accretors, and they found that $\xi \approx 0.49\pm 0.07$ and $L_{\rm prop} \sim10^{37}$\,erg\,s$^{-1}$ for \src. Here, $\xi=0.5$, $P\simeq0.467$\,s, $B=5\times10^{11}$\,G, and $\mu_{30}=BR^3/2\simeq0.32$ was assumed.  
To match the new observed propeller luminosity, one needs to significantly reduce the effective magnetosphere size, that is, the assumed magnetic field or value of $\xi$. In particular for $R_6=1.2$ \citep{2017MNRAS.466..906S}, a field of $\sim10^{10}$\,G or $\xi\sim0.06$ would need to be assumed.
Of course, the two parameters are correlated, so any combination is possible and quoted values only illustrate extreme values.

\begin{figure}[ht]
        \centering
                \includegraphics[width=\columnwidth]{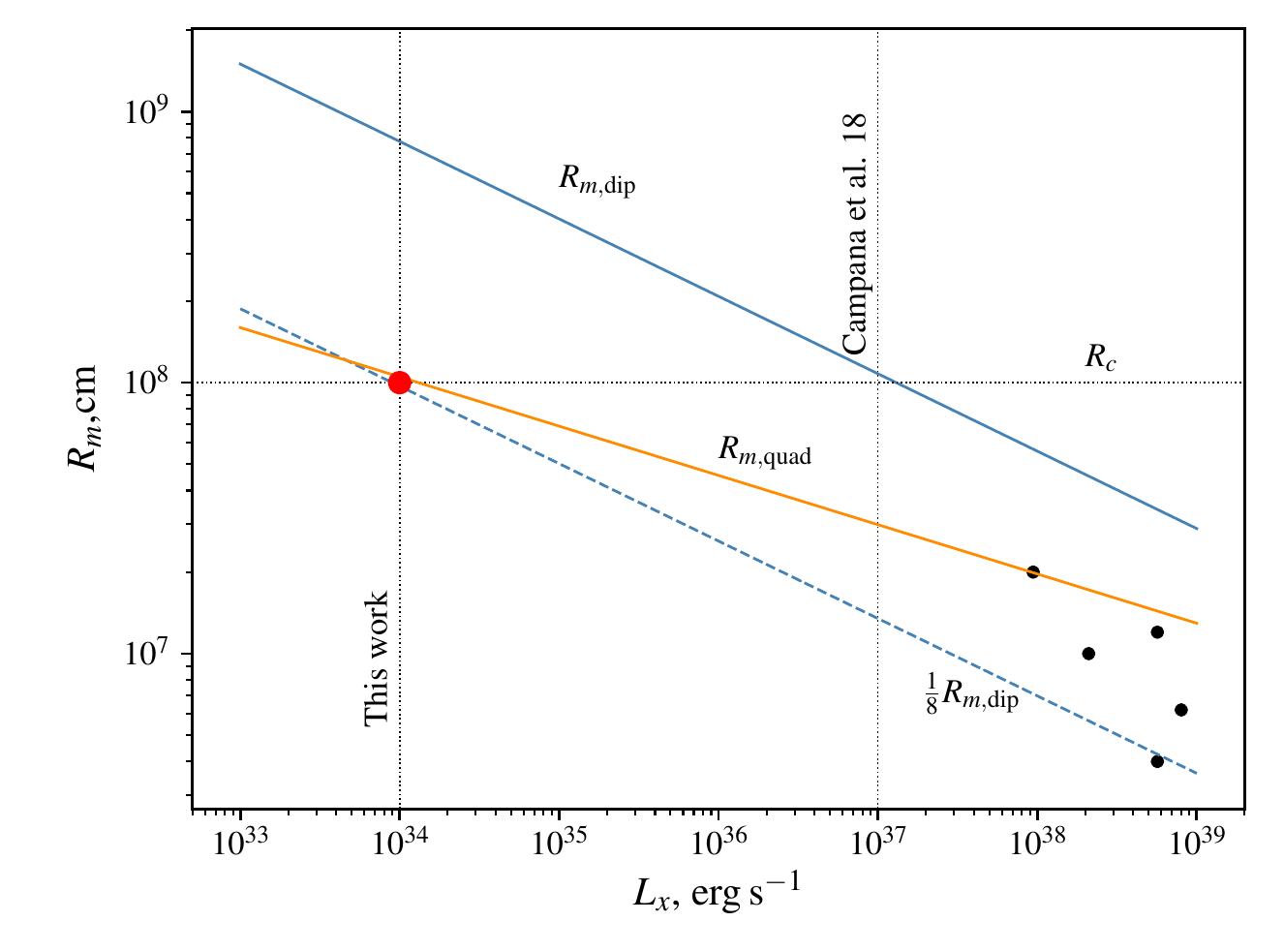}
        \caption{Estimates of the magnetospheric radius published in the literature (black points, see text for references) and obtained in this work (red circle) as a function of luminosity. Theoretical estimates of the magnetospheric radii for dipole $R_{m{,\rm dip}}$, assuming $\xi=0.49$ and $\mu_{30}=0.43$ \citep{2018A&A...610A..46C}, and  quadrupole $R_{m{,\rm quad}}$ with $\xi=0.8$ and $\mu_{30}=0.43$ \citep{2019A&A...626A.106M}, corotation radius $R_c$ and estimated luminosities for the propeller transition obtained in this work and by \cite{2018A&A...610A..46C} (black dotted lines) are also shown.}
        \label{fig:rin}
\end{figure}

Interestingly, the same issue has
already been identified and discussed in the context of source properties during an
outburst. In particular, \cite{2019A&A...626A.106M} conclude that the
magnetosphere must be small and estimated $\xi\sim0.09$ based on the observed
break frequency in the power density spectrum of the source close to the peak of
the outburst. As discussed by \cite{2019A&A...626A.106M}, modeling of the broad
iron line properties \citep{1999ApJ...517..436N,2014ApJ...796L...9D} and of the
soft blackbody component observed in the X-ray spectrum of the source
\citep{2015MNRAS.449.4288D,2015ApJ...804...43Y} and spin evolution of the source \citep{2017MNRAS.469....2S} also yield comparable estimates
for the inner accretion disk's radius (assuming a distance of 8\,kpc).
To resolve this issue, \cite{2019A&A...626A.106M}
suggest that the quadrupole field may dominate the interaction of the accretion disk. The magnetospheric radius
\begin{equation}
R_{m,{\rm quad}}\simeq2.4\times10^7\xi M^{1/11}R_6^{14/11}B_{11}^{4/11}L_{36}^{-2/11}\,\,\,{\rm cm},
\label{eq:rmquad}
\end{equation}
in this case, appears to be consistent with the estimates reported in the literature \citep{1999ApJ...517..436N,2014ApJ...796L...9D,2015MNRAS.449.4288D,2015ApJ...804...43Y} and the observed cyclotron line energy for $\xi\simeq1$ \citep{2019A&A...626A.106M}. As illustrated in Fig.~\ref{fig:rin}, it also turns out to be consistent with the propeller transitional luminosity derived in this paper if $R_6=1.2$, $\mu_{30}=0.43$, and $\xi=0.8$ are assumed, that is, a small estimated magnetosphere size could be explained if the magnetic field is dominated by a quadruple component.

On the other hand, simply scaling the estimate for the dipole field by factor $\sim8$, that is, reducing the assumed field to $\sim10^{10}$\,G or the effective magnetosphere size to $\xi\sim0.06$, gives an equally good (or bad) agreement. 
We note, however, that assuming a lower field is not only at odds with the observed cyclotron line energy, but it also makes it hard to explain why type I bursts are not observed \citep{1997ApJ...477..897B}. On the other hand, \cite{2018A&A...610A..46C} discuss that $\xi\sim0.5$ appears to be common for most magnetic accretors, so it is hard to explain why it must be so peculiarly low for \src, which makes a quadrupolar field hypothesis more appealing. 

Last but not least, it is important to emphasize that assuming a different distance to the source does not resolve the problem. Indeed, for a smaller distance, the propeller luminosity becomes even lower thereby increasing the discrepancy with the expectation given by Eq.~\ref{eq:lprop}. A lower assumed distance also implies a lower accretion rate and thus larger expected magnetosphere size in outburst, thereby increasing the discrepancy with the small observed value. On the other hand, increasing the assumed distance hardly helps to account for a difference of three orders of magnitude between the expected and observed propeller luminosities. A similar argument could be made for possible geometric beaming increasing the apparent luminosity of the source and this is thus equivalent to the situation when the distance to the source is over-estimated, as outlined above.

\section{Summary and conclusions}
Based on the deep \emph{XMM-Newton} observation of \src in quiescence and by revisiting the available archival observations, several results discussed above have been obtained:
\begin{itemize}
    \item The improved X-ray localization of the source allowed us to confirm the previously suggested optical counterpart as a G/K~III star at a distance consistent with the location of \src in the Galactic center. This identification also implies that the binary system is viewed nearly face on.
    \item A strong variability around $\sim10^{34}$\,erg\,s$^{-1}$, which was  revealed by \textit{XMM} (for the assumed distance of 8\,kpc), implies that the source definitively continues to accrete.
    \item Archival data suggest that the object is detected also at much lower (by one to two orders of magnitude) fluxes, but not intermediate flux levels. The observed bimodal flux distribution in the historical light curve and softening at the lowest flux levels thus strongly suggest that \src transitions to the propeller regime at around $\sim10^{34}$\,erg\,s$^{-1}$.
    \item The propeller transition takes place, therefore, at a much lower luminosity than previously assumed, which implies that the magnetosphere in \src is much more compact than can be expected for the field strength implied by the observed cyclotron line energy.
        
\end{itemize}
We emphasize that accretion seems to be the only plausible explanation for the observed variability of the source in quiescence, and thus our final conclusion regarding the absolute magnetosphere size is robust. 
It is also in line with several independent estimates of the magnetosphere size during the outburst, which led to the same conclusion, that is, that the magnetosphere is an order of magnitude smaller than could be expected for the field strength deduced based on the observed cyclotron line energy. 

The options to explain this finding have already been discussed in the literature and are fairly limited: Besides potential possible peculiarities in the interaction of the radiatively-dominated accretion disk with the magnetosphere,
a grossly overestimated magnetic field, and the presence of a strong multipole field component were suggested. 
Based on the fact that the magnetosphere also remains compact in quiescence, we conclude that the first option can be ruled out since the accretion disk is definitively not in a radiative-pressure dominated state in quiescence. Considering that the cyclotron line was independently discovered by several authors using the data from several instruments obtained in two independent outbursts, and the absence of Type I bursts expected for such a low field, the second option also appears unlikely. We conclude, therefore, that \src is a prime candidate neutron star with  a complex field morphology.
This hypothesis can be tested, for instance, through polarimetric observations with the upcoming \textit{IXPE} \citep{2016SPIE.9905E..17W} and \textit{eXTP} \citep{2016SPIE.9905E..1QZ}.

\section*{Acknowledgements}
We thank the anonymous referee for very useful comments which helped to substantially improve the manuscript.
This research has made use of data obtained from the Chandra Source Catalog, provided by the Chandra X-ray Center (CXC) as part of the Chandra Data Archive.
The authors thank the Russian Science Foundation (grant 19-12-00423) for financial support. VS also thanks Deutsche  Forschungsgemeinschaft  (DFG) for financial support (grant WE 1312/51-1).
We also acknowledge the support from the Academy of Finland travel grants 331951 (ST, JM), the Vilho, Yrj\"o and Kalle V\"ais\"al\"a Foundation (ST, JM),  the German Academic Exchange Service (DAAD) travel grant 57525212 (VD, VS, LJ). JL appreciates the support from the National Natural Science Foundation of China under Grants 11733009 and U1938103.

\bibliographystyle{aa}
\bibliography{biblio}
\end{document}